\documentclass[12pt,preprint]{aastex}

\usepackage{graphicx}

\newcommand{\dd}{\mathrm{d}}
\newcommand{\ee}{\mathrm{e}}

\newcommand{\eqref}[1]{(\ref{#1})}

\shorttitle{Inelastic Collisions in General Relativity}
\shortauthors{Hennig et al.}

\begin{document}

\title{Thermodynamic Description of Inelastic Collisions in General
Relativity}

\author{J\"org Hennig, Gernot Neugebauer}
\affil{Theoretisch-Physikalisches Institut,
Friedrich-Schiller-Universit\"at
Jena, Max-Wien-Platz 1, D-07743 Jena, Germany}
\email{J.Hennig@tpi.uni-jena.de}
\and
\author{Marcus Ansorg}
\affil{Max-Planck-Institut f\"ur Gravitationsphysik,
Albert-Einstein-Institut, Am M\"uhlenberg 1,  D-14476 Golm, Germany}
\email{marcus.ansorg@aei.mpg.de}

%\author{J\"org Hennig\altaffilmark{1}, Gernot Neugebauer\altaffilmark{1}
%and Marcus Ansorg\altaffilmark{2}}
%\affil{\altaffilmark{1}Theoretisch-Physikalisches Institut,
%Friedrich-Schiller-Universit\"at
%Jena,\\ Max-Wien-Platz 1, D-07743 Jena, Germany}
%\email{J.Hennig@tpi.uni-jena.de}
%\affil{\altaffilmark{2}Max-Planck-Institut f\"ur Gravitationsphysik,
%Albert-Einstein-Institut,\\ Am M\"uhlenberg 1,  D-14476 Golm, Germany}
%\email{marcus.ansorg@aei.mpg.de}

%%%%%%%%%%%%%%%%%%%%%%%%%%%%%%%%%%%%%%%%%%%%%%%%%%%%%%%%%%%%%%%%%%%%%%%%%%%%
\begin{abstract}
We discuss head-on collisions of neutron stars and disks of dust (\lq\lq
galaxies\rq\rq) following the ideas of equilibrium thermodynamics, which
compares equilibrium states and avoids the description of the dynamical
transition processes between them. As an always present damping mechanism,
gravitational emission results in final equilibrium states after the
collision. In this paper we calculate selected final configurations
from initial data of colliding stars and disks by making use
of conservation laws and solving the Einstein equations. Comparing
initial and final states, we can decide for which initial parameters two
colliding neutron stars (non-rotating Fermi gas models) merge into a
single neutron star and two rigidly rotating disks form again a final
(differentially rotating) disk of dust. For the neutron star collision
we find a maximal energy loss due to outgoing gravitational radiation of
$2.3\%$ of the initial mass while the corresponding efficiency for
colliding disks has the 
much larger limit of $23.8\%$.
\end{abstract}


\maketitle

%%%%%%%%%%%%%%%%%%%%%%%%%%%%%%%%%%%%%%%%%%%%%%%%%%%%%%%%%%%%%%%%%%%%%%%%%%%%%
\section{Introduction}

Collisions of compact objects are an important source of gravitational
radiation. 
Much effort has recently been made to develop numerical methods and
codes describing and simulating the underlying hydrodynamical and
gravitational phenomena.
After the pioneering work on numerical black hole
evolutions
by Eppley and Smarr in the
1970's
[see e.g.
%\cite{Eppley,Smarr}
Eppley (1975) and Smarr et al. (1976)], head-on collisions were re-investigated
in the 1990's
%\cite{Anninos1, Anninos2, Anninos3}
(Anninos et al. 1993, 1995, 1998)
with good agreement
between numerical and perturbation-theoretical results. Long-term-stable
evolutions of black hole and neutron star collisions were
successfully performed in the last two years
%\cite{Sperhake1, Fiske, Zlochower,Sperhake2, Loeffler}.
(Sperhake et al. 2005; Fiske et al. 2005; Zlochower et al. 2005;
Sperhake 2006; L\"offler et al. 2006).

\begin{figure}\centering
\includegraphics[scale=0.3]{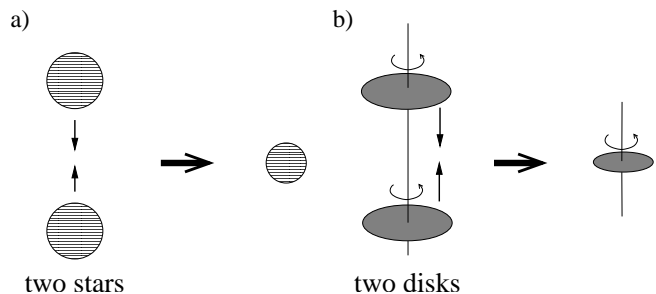} %frueher: model.eps
\caption{Model: collisions of spherically symmetric stars or rigidly
rotating disks of dust}
\label{fig_model}
\end{figure}

From a mathematical point of view collision processes are typical
examples for initial-boundary problems. In particular,
we will discuss head-on
collisions of spheres and disks\footnote{Disk-like
matter configurations play an
important role in astrophysics, e.g. as models for galaxies,
accretion disks or  intermediate phases in the merger process
of two neutron stars.}, see
Fig.~\ref{fig_model}.
Starting with bodies separated by a large (\lq\lq
infinite\rq\rq) distance we may model the initial situation by a
quasi-equilibrium configuration of two isolated bodies.
Corresponding solutions for spheres and
(rigidly rotating) disks can be found in the literature, see e.g.
%\cite{MTW,Shapiro,Neugebauer1,Neugebauer2,Neugebauer3}
Misner et~al. (2002), Shapiro \& Teukolsky (1983) and Neugebauer \& Meinel
(1993, 1994, 1995).
The dynamical
phase of the collision process is always accompanied by gravitational
radiation. This damping mechanism results again in the formation of an
equilibrium configuration after the collision.
The rigorous mathematical description of the dynamical transition phase
is difficult and requires extensive numerical investigations. However,
interesting information about the collision can be obtained by comparing
the initial and final (equilibrium) states. This thermodynamic idea
avoids the analysis of the transition process
and reduces the mathematical effort to solving the
Einstein equations for the end products, which are stationary and
axisymmetric in our case. The
solution makes use of conservation laws which transfer
data extracted from the initial configurations (spheres and disks
before the collision) to the final configurations.

While the \emph{initial} configurations are available
the calculation of the \emph{final} states is rather difficult.
To cope with this problem for head-on colliding stars and disks,
we will make use of two
heuristic principles:
\begin{itemize}
   \item[1)] Perfect fluid configurations at rest are
              spherically symmetric. Hence, the end product of two head-on
              colliding spheres without angular momentum  is again a
              sphere
              (a fluid ball or a
              Schwarzschild black hole).
   \item[2)] Dust configurations
              are two-dimensional
              (\lq\lq extremely flattened\rq\rq) and axisymmetric
              (with non-vanishing angular momentum). Consequently,
              the dust matter after a head-on
              collision of two disks of dust is again two-dimensional and
              axisymmetric (a compact disk, a disk surrounded by
              dust rings or a black hole surrounded by dust rings).
\end{itemize}
Though plausible, these principles have not been proved rigorously so
far\footnote{In this context we refer to an new approach by
Masood-ul-Alam (2007).}. For proofs under special assumptions see
%\cite{Beig, Lindblom}.
Beig \& Simon (1992) and Lindblom \& Masood-ul-Alam (1994). 

As illustrated in Fig.~\ref{fig_model} we will confine ourselves to two problems:
\begin{itemize}
   \item[a)] head-on collisions of two identical spheres (stars) merging
             into a single fluid ball and
   \item[b)] head-on collisions of two identical disks of dust
             (galaxies) merging into a single disk.
\end{itemize}

We will be able to formulate \emph{necessary} conditions for the
formation of these balls or disks. Obviously, the conditions will
restrict the parameters of the initial configuration; a violation of the
conditions would necessarily lead to other final states such as to
black holes or central disks surrounded by rings. To express the
parameters of the admissible initial parameters --- the first goal of
this paper --- we have to solve the Einstein equations (numerically but)
rigorously and to make use of the conservation laws  for baryonic mass
and angular momentum. There is no obstacle to an extension of the
method. One could start a systematic investigation of other possible
final states after the collision (black holes, black holes with rings
etc.) making use of symmetries, conservation laws and the heuristic
principles 1) and 2). An important point of the procedure would be the
stability analysis of the end products. As for our investigation, there
is important evidence from Newtonian gravity that rigidly or
differentially rotating disks of dust are unstable. Nevertheless we can
expect that \lq\lq stabilizing\rq\rq effects (pressure due to internal
kinetic energy) do not falsify our other goal --- to estimate the
maximal contribution of gravitational
radiation to the total energy loss $\Delta
M$. In general, the total energy loss calculated via the
comparison of initial and final equilibrium configurations is only an
upper limit for the energy loss (efficiency) due to gravitational
emission. (It includes energy loss due to non-gravitative radiation or
mass ejection during the collision). We will present a disk collision
model which is exclusively damped by gravitational radiation. The
resulting differentially rotating disk will be compared with a rigidly
rotating disk of the same baryonic mass and angular momentum formed from
the same initial disks under the additional influence of dissipative
processes in the matter (see \ref{DR}).
Thus we can  compare the efficiencies of
the two forms of dissipation.

In Sec.~\ref{Stars} we discuss, as introductory examples,  
the merger of
two Schwarzschild stars and the collision of
two (Fermi gas)
neutron stars.
Sec.~\ref{Disks} contains the main part of this paper which is
dedicated to the investigation of disk collisions.
These discussions are based on a novel solution of the Einstein
equations describing the final configuration.
Here we continue the analysis of
a previous paper \cite{Hennig},
in which we discussed the collisions of rigidly rotating disks of
dust with parallel (or antiparallel) angular momenta under the simplifying
assumption that the final disk be again a \emph{rigidly} rotating (or
rigidly counterrotating) disk of
dust.
This assumption can only be justified if friction processes between the
disk rings provide for a constant angular velocity throughout the disk.
This model seems to be somewhat artificial and unsuited to determining the
contribution of gravitational radiation to the total energy
loss. Interestingly, our present investigation will show that the
frictional contribution to the total energy loss for colliding rigidly
rotating disks is comparably small. 
%%%%%%%%%%%%%%%%%%%%%%%%%%%%%%%%%%%%%%%%%%%%%%%%%%%%%%%%%%%%%%%%%%%%%%%%%%%%
\section{Star collisions}\label{Stars}

%%%%%%%%%%%%%%%%%%%%%%%%%%%%%%%%%%%%%%%%%%%%%%%%%%%%%
\subsection{Introductory example: Schwarzschild stars}
In order to demonstrate the method, we study the collision 
of two Schwarzschild
stars, i.e. spherically symmetric perfect fluid stars with a constant
mass density, $\mu=\mathrm{constant}$.
Though not very realistic, this model illustrates the main steps
of the method.

The matter of a Schwarzschild star is described by the perfect fluid
energy-momentum tensor
\begin{equation}\label{2}
T^{ij}=(\mu+p)u^i u^j+p g^{ij}
\end{equation}
with the pressure
\begin{equation}\label{2a}
p(r)=\frac{\sqrt{1-\frac{8\pi\mu}{3}r^2}
-\sqrt{1-\frac{8\pi\mu}{3}r_0^2}}{3\sqrt{1-\frac{8\pi\mu}{3}r_0^2}
-\sqrt{1-\frac{8\pi\mu}{3}r^2}}\mu,
\end{equation}
where $u^i$, $r$ and $r_0$ are the four-velocity,
the radial coordinate and the coordinate radius of
the star, respectively.
The interior Schwarzschild metric can be
written as 
\begin{equation}\label{1}
\dd s^2=
\frac{\dd r^2}{1-\frac{8\pi\mu}{3}r^2}
+r^2(\dd\vartheta^2+\sin^2\vartheta\,\dd\varphi^2)
-\left(\frac{3}{2}\sqrt{1-\frac{8\pi\mu}{3}r_0^2}
-\frac{1}{2}\sqrt{1-\frac{8\pi\mu}{3}r^2}\right)\dd t^2,
\end{equation}
and  the
exterior Schwarzschild solution is
\begin{equation}\label{2aa}
\dd s^2=\frac{\dd
r^2}{1-\frac{2M}{r}}
+r^2\left(\dd\vartheta^2+\sin^2\vartheta \dd\varphi^2\right)
-\left(1-\frac{2M}{r}\right)\dd t^2.
\end{equation}
Note that we use the normalized units where $c=1$ for the speed of
light and $G=1$ for Newton's gravitational constant.

The gravitational mass $M$,
\begin{equation}\label{3}
M=\frac{4\pi\mu}{3}r_0^3,
\end{equation}
follows from the matching condition at the star's surface and the
baryonic mass $M_0$ is given by
\begin{equation}\label{4}
M_0=\int\limits_{t=t_0}\mu u^t\sqrt{-g}\,\dd r\dd\vartheta\dd\varphi
=4\pi\mu\int\limits_0^{r_0}\frac{r^2\dd
r}{\sqrt{1-\frac{8\pi\mu}{3}r^2}}.
\end{equation}

Now we apply these formulae to the head-on collision of two stars. 
Restricting ourselves to collisions of two identical 
Schwarzschild stars we assume, that the final star
be again a Schwarzschild star and have the same mass density
(e.g. nuclear matter density),
\begin{equation}\label{4a}
\tilde\mu=\mu,
\end{equation}
where from now on tildes denote quantities after the collision.

The conservation of baryonic mass during the collision
process
%\footnote{In all our considerations in this paper
%we will exclude mass loss by
%matter ejection  or by radiation from
%thermonuclear reactions during the collision process.}
\begin{equation}\label{5}
\tilde M_0=2M_0,
\end{equation}
allows one to calculate the parameters of the final star as a function of
the initial parameters. With \eqref{3}, \eqref{4} and \eqref{4a} the
conservation equation \eqref{5} can be written as
\begin{equation}\label{6}
\arcsin\left(\sqrt{\frac{2M}{r_0}}\frac{\tilde r_0}{r_0}\right)
-\sqrt{\frac{2M}{r_0}}\frac{\tilde r_0}{r_0}
\sqrt{1-\frac{2M}{r_0}\frac{\tilde r_0^2}{r_0^2}}
=
2\left(\arcsin\sqrt{\frac{2M}{r_0}}
-\sqrt{\frac{2M}{r_0}}
\sqrt{1-\frac{2M}{r_0}}\right),
\end{equation}
i.e.  the radius ratio $\tilde r_0/r_0$ is a function of the
initial mass-radius ratio
$2M/r_0$. Hence, we may express the efficiency $\eta$
of conversion of mass into gravitational radiation,
\begin{equation}\label{7}
\eta=1-\frac{\tilde M}{2M}=1-\frac{1}{2}\left(\frac{\tilde r_0}{r_0}\right)^3,
\end{equation}
and the mass-radius ratio of the final star,
\begin{equation}\label{8}
\frac{2\tilde M}{\tilde r_0}=\frac{2M}{r_0}\left(\frac{\tilde r_0}{r_0}\right)^2,
\end{equation}
in terms of $2M/r_0$.

\begin{figure}\centering
\includegraphics[scale=0.96]{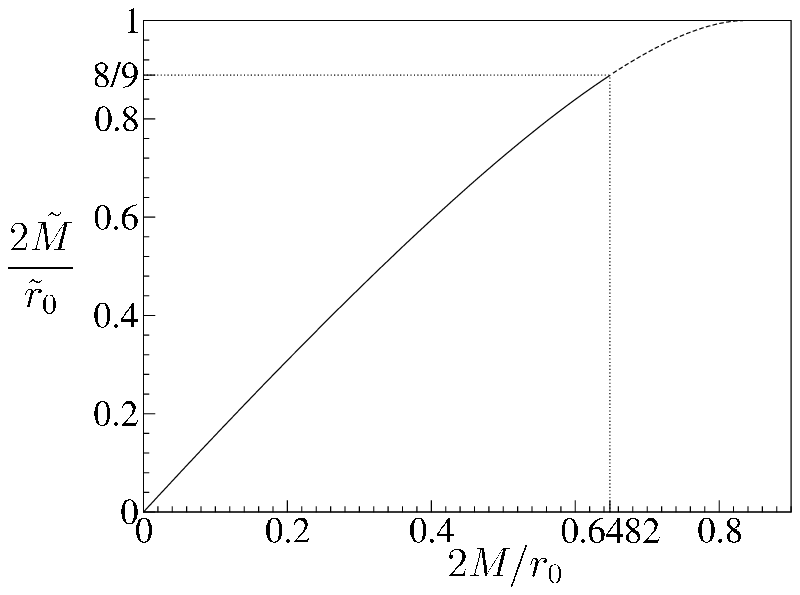}
\includegraphics[scale=0.96]{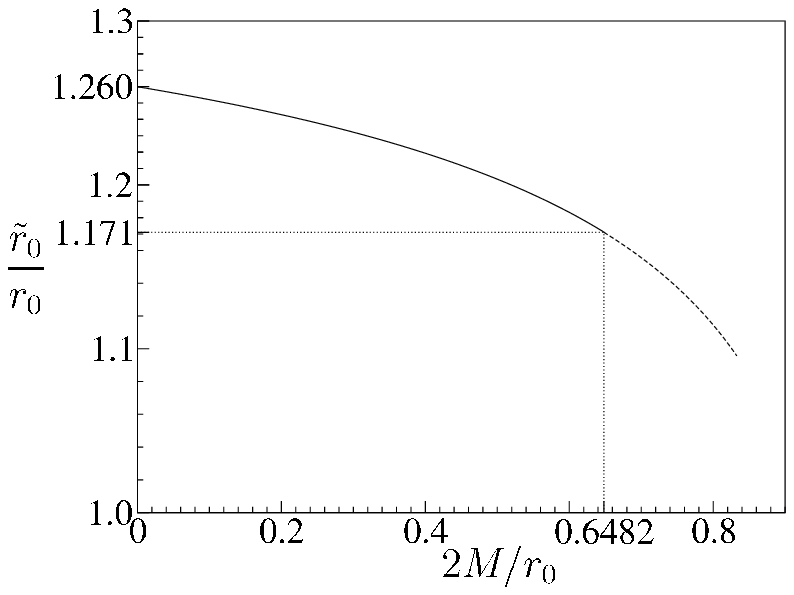}\\[2ex]
\includegraphics[scale=0.96]{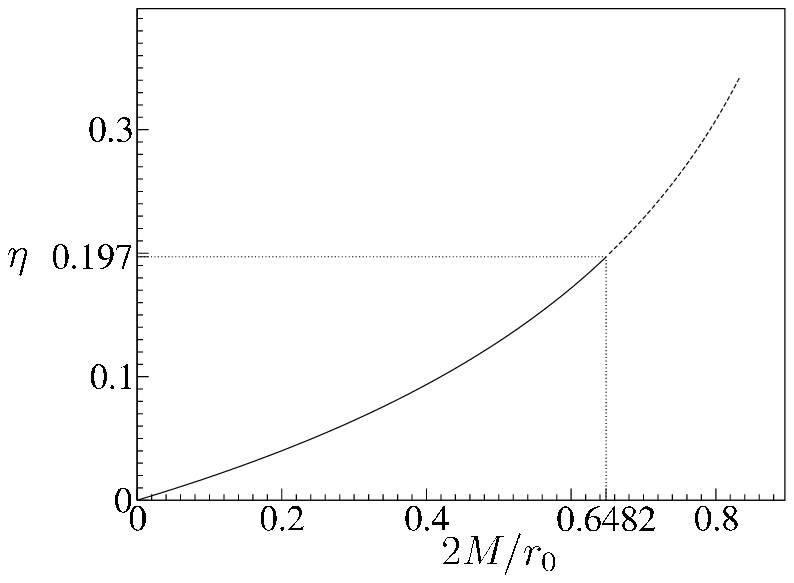}
\caption{Parameter relations for colliding Schwarzschild stars:
The final mass-radius ratio $2\tilde M/\tilde r_0$, the
radius ratio $\tilde r_0/\tilde r$ and the efficiency
$\eta$ are plotted as functions of the initial
mass-radius ratio $2M/r_0$.
Dashed parts of the curves mark regions inaccessible due to the Buchdahl
inequality $2\tilde M/\tilde r_0< 8/9$.}
\label{fig_Schwarzschild}
\end{figure}

The resulting parameter relations are plotted in
Fig.~\ref{fig_Schwarzschild}. For Schwarzschild stars the coordinate
radius is restricted by the Buchdahl condition, i.e.
\begin{equation}\label{8a}
r_0>\frac{9}{8}\times 2M,
\quad
\tilde r_0>\frac{9}{8}\times 2\tilde M.
\end{equation}
As a consequence, the first plot
shows, that \lq\lq relativistic\rq\rq\ initial stars with
$2M/r_0>0.6482\dots$ can never merge into a new Schwarzschild star with the
same mass density $\mu$. The \lq\lq physical\rq\rq\ parts of the
parameter relations are shown as solid curves while the forbidden parts
are dashed.
According to the third plot the efficiency $\eta$ cannot exceed a
maximal value of $\eta_\textrm{max}\approx 19.7\%$.
 
%%%%%%%%%%%%%%%%%%%%%%%%%%%%%%%%%%%%%%%%%%%%%%%%%%%%%%%%%%%
\subsection{Neutron stars: Completely degenerate ideal Fermi gas}

In order to extend the discussion of the previous section
to a more realistic star model, we replace the equation of state
$\mu=\textrm{constant}$ by the equation  for  a completely
degenerate ideal Fermi gas of neutrons.

The (interior) line element of a spherically symmetric star can be
written as 
\begin{equation}\label{9}
\dd s^2=
\ee^{2\lambda(r)}\dd r^2
+r^2(\dd\vartheta^2+\sin^2\vartheta\,\dd\varphi^2)
-\ee^{2\nu(r)}\dd t^2
\end{equation}
and the matter is again described by the perfect fluid energy-momentum tensor
\begin{equation}\label{10}
T^{ij}=(\mu+p)u^i u^j+p g^{ij}.
\end{equation}
With the definition of a new metric function $m(r)$ by
\begin{equation}\label{11}
\ee^{2\lambda(r)}=\frac{1}{1-\frac{2m(r)}{r}},
\end{equation}
the field equations can be written in the TOV form,
%\cite{Shapiro}
see e.g. Shapiro \& Teukolsky (1983),
\begin{equation}\label{12}
\frac{\dd m}{\dd r}=4\pi r^2\mu, \quad m(0)=0
\end{equation}
\begin{equation}\label{13}
\frac{\dd p}{\dd r}=-\frac{m}{r^2}\mu\left(1+\frac{p}{\mu}\right)
\left(1+\frac{4\pi r^3 p}{m}\right)\left(1-\frac{2m}{r}\right)^{-1},
\quad p(0)=p_\textrm{c}
\end{equation}
\begin{equation}\label{14}
\frac{\dd \nu}{\dd r} = -\frac{1}{\mu}\frac{\dd p}{\dd r}
\left(1+\frac{p}{\mu}\right)^{-1},
\end{equation}
where $p_\textrm{c}$ is the pressure in the center of the star.

\begin{figure}\centering
\includegraphics[scale=0.98]{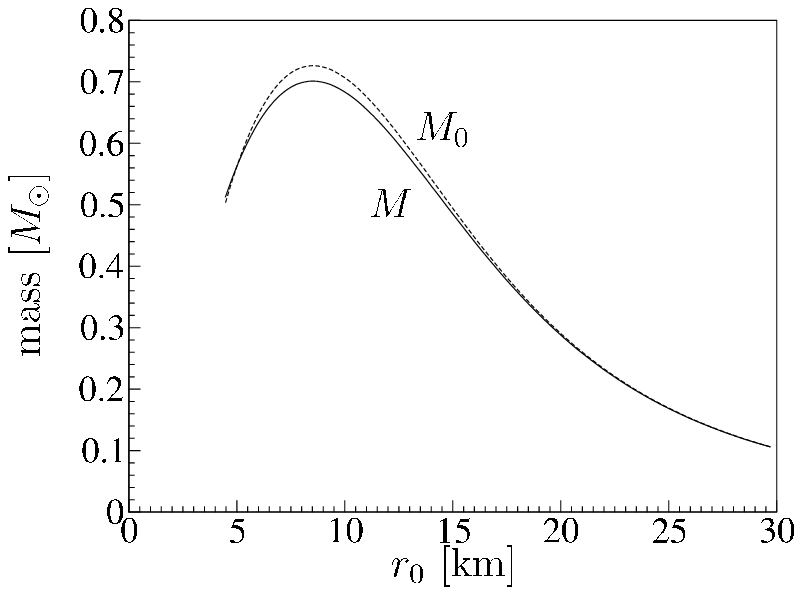}
\includegraphics[scale=0.98]{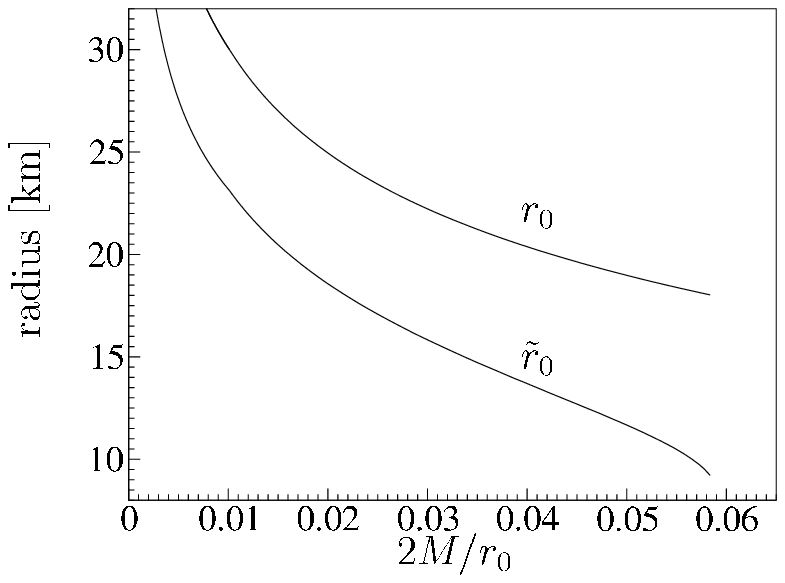}\\[2ex]
\includegraphics[scale=0.98]{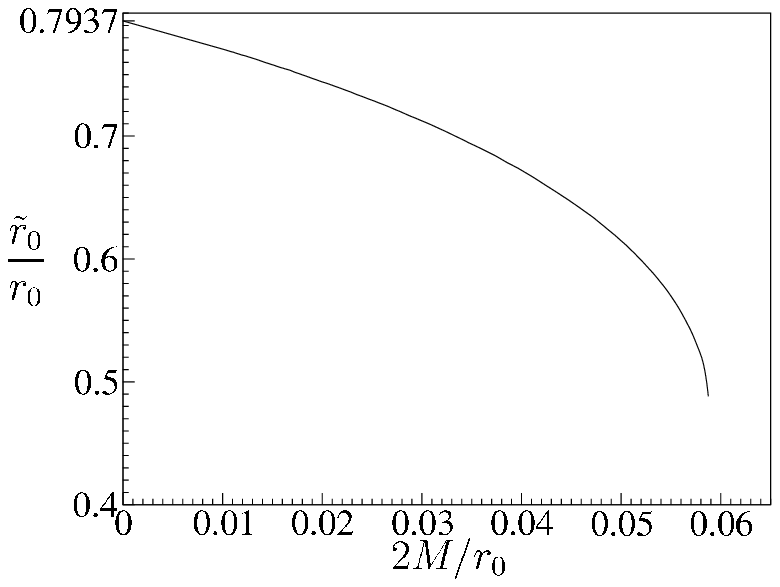}
\includegraphics[scale=0.98]{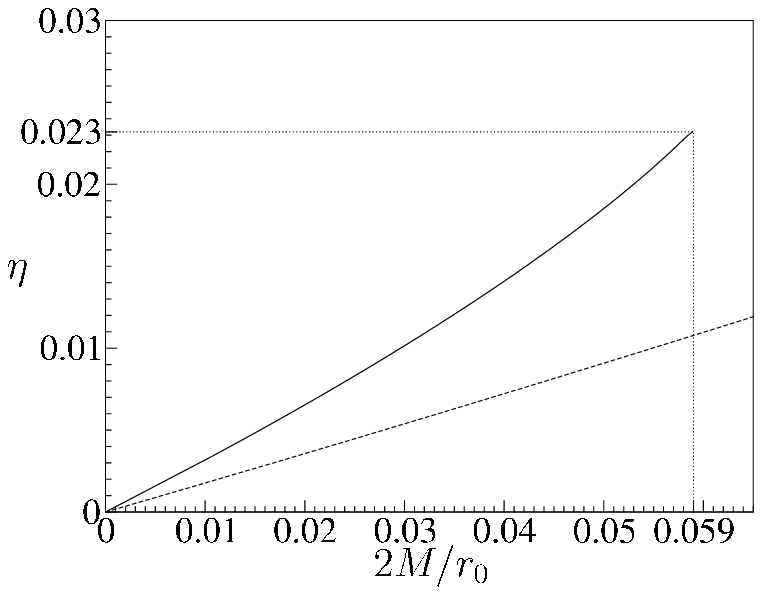}
%frueher: Neutronenstern_d.eps
\caption{Parameter relations for the collisions of neutron stars made up
of degenerate neutrons [cf. \eqref{15}]. First
plot: mass-radius relations for the baryonic mass $M_0$ and the
gravitational mass $M$. Second plot: initial radius $r_0$ and final
radius $\tilde r_0$ as functions of the mass-radius ratio $2M/r_0$.
Third plot: change of the coordinate
radius. Fourth plot: efficiency $\eta$ compared to the efficiency of
the collision of Schwarzschild stars (dashed curve). }
\label{fig_Neutronenstern}
\end{figure}

We will solve
these equations for the
completely degenerate ideal fermi gas of neutrons
with the equation of state
%\cite{Shapiro}
[see e.g. Shapiro \& Teukolsky (1983)]
\begin{equation}\label{15}
p=c_1f(x),\quad \rho=c_2 x^3, \quad \mu=\rho+c_1 g(x),
\end{equation}
where
\begin{equation}\label{16}
f(x)=x(2x^2-3)\sqrt{1+x^2}+3\ln(x+\sqrt{1+x^2}),
\end{equation}
\begin{equation}\label{17}
g(x)=8x^3(\sqrt{1+x^2}-1)-f(x),
\end{equation}
\begin{equation}\label{18}
c_1=\frac{\pi m_\textrm{n}^4}{3h^3}, \quad c_2=\frac{8\pi m_\textrm{n}^4}{3h^3}
\end{equation}
with the neutron mass $m_\textrm{n}=1.6749286\times 10^{27}\,\mathrm{kg}$
and Planck's constant $h=6.626076\times 10^{-34}\,\mathrm{Js}$.
By
solving the TOV equations \eqref{12} and \eqref{13} with the
equation of state \eqref{15} for a sequence of values of the central density
one can calculate the corresponding radii of the stars as the first zero
$r_0$ of $p(r)$, their gravitational mass from $M=m(r_0)$,
and their  baryonic mass as
\begin{equation}\label{19}
M_0=4\pi\int\limits_0^{r_0}\frac{\rho(r)r^2\dd r}{\sqrt{1-\frac{2m(r)}{r}}}.
\end{equation}
The resulting mass-radius relations are shown in the first plot of
Fig.~\ref{fig_Neutronenstern}.

Again the baryonic mass is an invariant of the collision, i.e.
\begin{equation}\label{20}
\tilde M_0=2M_0
\end{equation}
for the collision of two identical initial stars.
This equation has to be analysed together with the mass-radius relations.
(Thereby we
take into account only stars in the monotonic decreasing part of the
mass-radius relation $M_0(r_0)$.) The resulting
parameter relations are shown in
the remaining plots of Fig.~\ref{fig_Neutronenstern}.
For the  maximum of the efficiency one finds
$\eta_\textrm{max}\approx 2.3\%$, i.e. a
comparably small value in view of the maximal efficiency
$\eta_\textrm{max}\approx 19.7\%$
for collisions of
Schwarzschild stars. The reason is the relatively small maximal mass of
$M_\textrm{max}\approx 0.7 M_\odot$
permitted by the equation of state \eqref{15} that excludes
highly relativistic values for the mass-radius ratio
$2M/r_0$. However, compared to Schwarzschild stars with the same parameter
$2M/r_0$ the collisions of neutron stars are more efficient,
cf. the last plot in
Fig.~\ref{fig_Neutronenstern}.

Another difference is the change of the coordinate radii.
While two Schwarzschild stars merge into a Schwarzschild star with a
coordinate radius bigger than the initial radius,
$\tilde r_0/r_0>1$ (cf. Fig.~\ref{fig_Schwarzschild}),
the resulting neutron star is smaller than the initial neutron stars,
$\tilde r_0/r_0<1$ (cf. Fig.~\ref{fig_Neutronenstern}).

%%%%%%%%%%%%%%%%%%%%%%%%%%%%%%%%%%%%%%%%%%%%%%%%%%%%%%%%%%%%%%%%%%%%%%%%%%%%%
\section{Disk collisions}\label{Disks} 

Collisions of disks of dust require more effort. In particular,
the discussion of the final equilibrium state is based on a solution
of a free boundary value problem to the Einstein equations.
At the first glance, the conservation laws for baryonic mass and angular
momentum are not sufficient to formulate a complete set of boundary
conditions for the configuration after the collision. However, excluding
non-gravitational dissipation, we may
replace the global conservation laws, as used in Sec.~\ref{Stars}, by
local ones.
Due to the geodesic motion of dust particles, the baryonic mass and the
angular momentum of each of the rings forming the disk are conserved
separately, see Fig.~\ref{fig_local_conservation}. Using such \emph{local}
conservation laws we will be able to solve
(numerically) the boundary value problem for the final state after the
head-on collision of two aligned rigidly rotating disks
of dust with parallel angular momenta, cf. Fig.~\ref{fig_model}.

%%%%%%%%%%%%%%%%%%%%%%%%%%%%%%%%%%%%%%%%

\subsection{Initial disks: Rigidly rotating disks of dust}

The free boundary value problem for the relativistic rigidly rotating
disk of dust (RR disk) was
discussed by Bardeen and Wagoner (1969, 1971) using approximation methods
and
analytically solved in terms of ultraelliptic theta functions
by Neugebauer and Meinel (1993, 1994, 1995)
using the Inverse Scattering
Method. 
The line element of the stationary (Killing vector: $\xi^i$) and
axisymmetric (Killing vector: $\eta^i$) space-time may be written
in the Weyl-Lewis-Papapetrou standard form
\begin{equation}\label{21}
\dd s^2=\ee^{-2U}[\ee^{2k}(\dd
\rho^2+\dd\zeta^2)+\rho^2\dd\varphi^2]-\ee^{2U}(\dd t+a\dd\varphi)^2,
\quad \xi^i=\delta^i_t,\quad \eta^i=\delta^i_\varphi,
\end{equation}
where the metric potentials
$U=U(\rho,\zeta)$, $k=k(\rho,\zeta)$ and $a=a(\rho,\zeta)$ are
given in terms of ultraelliptic theta functions.

The matter of the disk of dust is described by the energy-momentum tensor
\begin{equation}\label{22}
T^{ij}=\varepsilon(\rho)\delta(\zeta)u^i u^j,
\end{equation}
where $\varepsilon(\rho)\delta(\zeta)$ is the mass density with
$\delta(\zeta)$ as Dirac's $\delta$-distribution. Due to the
symmetries, the
four-velocity of the dust particles is a linear combination of the two
killing vectors,
\begin{equation}\label{22a}
u^i=\ee^{-V_0}(\xi^i+\Omega_0\eta^i),\quad u^i u_i=-1,
\end{equation}
whence
\begin{equation}\label{22b}
(\xi^i+\Omega_0\eta^i)(\xi_i+\Omega_0\eta_i)=-\ee^{2V_0},
\end{equation}
where $\Omega_0$ is the angular velocity of the particles forming the
disk and $V_0$ is a redshift parameter. Rigid rotation means
$\Omega_0=\textrm{constant}$ in the disk.
Since dust particles move geodesically
this assumption implies $V_0=\textrm{constant}$ in the disk. Hence, the boundary
condition \eqref{22b} and as a consequence the RR disk solution contains
two constant parameters.
Alternatively to $\Omega_0$ and $V_0$, one may choose the
coordinate radius $\rho_0$ of the disk and a centrifugal parameter
$\mu=2\Omega_0^2\rho_0^2\ee^{-2V_0}$ [$\mu\to 0$ turns out to be the
Newtonian limit and $\mu\to 4.62966\dots$ the ultrarelativistic limit,
where the disk approaches the extreme Kerr black hole,
cf.
%\cite{Neugebauer2} and \cite{Neugebauer4}
Neugebauer \& Meinel (1994) and Neugebauer et al. (1996) 
for these
and further properties].
%%%%%%%%%%%%%%%%%%%%%%%%%%%%%%%%%%%%%%%%%%%%%
\subsection{Final disk: Differentially rotating disk of dust}
\label{DR}

In a previous paper \cite{Hennig} we discussed head-on collisions
of two (identical) rigidly rotating disks of dust merging into one
\emph{rigidly} rotating disk of dust. The model excluded mass ejection
and made use of the conservation of baryonic mass and angular momentum
(axisymmetry). From a thermodynamic point of view rigid rotation of the
final disk means thermodynamic equilibrium, which is a result of 
dissipative processes during the dynamical phase. Hence, the energy
difference between the initial state (two separated disks) and the final
state (one rigidly rotating disk) is influenced by irreversible
processes in the matter and outgoing electromagnetic radiation as well
as by emission of gravitational waves. The intention of this paper is to
compare the contribution of these two effects by calculating the end
product of a purely gravitational collision process which we expect to
be a \emph{differentially} rotating disk of dust. Note that our
thermodynamic analysis enables us to formulate \emph{necessary}
conditions for the parameters of the initial disks ($\mu$ restricted) to
permit the formation of a final \emph{disk}. To obtain \emph{sufficient}
conditions one would have to solve the Einstein equations for the
time-dependent transition phase, which is outside the scope of this paper.

In the next subsection we will give a brief summary of the previous
paper. 
After that, we will see that the \emph{local} conservation of baryonic mass
and angular
momentum is sufficient to calculate the final
differentially rotating disk (numerically).
Differentially rotating disks with arbitrary rotation law 
have already been studied \cite{Ansorg1, Ansorg2}.
The point made here is that we are able to formulate a physically
motivated rotation law as a result of a collision process.

%%%%%%%%%%%%%%%%%%%%%%%%%%%%%%%%%%%%%%%%%%%%%
\subsubsection{Formation of rigidly rotating disks}
\label{AssumptionRR}

\begin{figure}\centering
\includegraphics[scale=1]{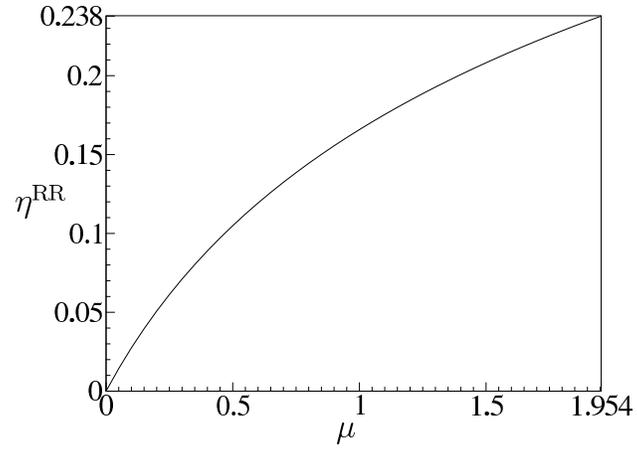}
\caption{The efficiency $\eta_{\textrm{RR}}$
for the formation of an RR disk from two initial
RR disks as a function of the centrifugal parameter $\mu$ of the initial
disks. $\eta^{\textrm{RR}}$ is an upper limit for the energy
loss due to gravitational
radiation.}
\label{fig_efficiency(RR)}
\end{figure}

For the formation of an RR disk from two colliding RR disks under the
influence of friction processes
the conservation equations for baryonic mass and angular momentum,
\begin{equation}\label{23}
\tilde M_0=2M_0,\quad \tilde J=2J,
\end{equation}
are sufficient to calculate the parameters of the final disks as functions
of the initial parameters. These equations and
explicit formulae connecting  the gravitational mass $M$,
the baryonic mass $M_0$ and the angular momentum $J$ of the RR disk
allowed us to calculate
the efficiency $\eta^{\textrm{RR}}=1-\tilde M/2M$
as a function of the initial centrifugal
parameter $\mu$, cf. Fig.~\ref{fig_efficiency(RR)}. It should be
emphasized once again,
that this efficiency measures the total energy loss including
friction. Therefore $\eta$ is only an \emph{upper limit} for the energy of the
gravitational emission. 
We obtained a maximal value of $\eta^{\textrm{RR}}_{\textrm{max}}\approx
23.8\%$
\cite{Hennig}.

Furthermore, it turned out that
the formation of RR disks from two colliding RR disks is only possible
for a rather restricted interval
$0<\mu<1.954\dots$ of the initial centrifugal parameter $\mu$.
 If $\mu$
exceeds this limit, the
collision must lead to other final states,
e.g. black holes or black holes surrounded by  matter rings.

%%%%%%%%%%%%%%%%%%%%%%%%%%%%%%%%%%%%%%%%%%%
\subsubsection{Local conservation equations}

\begin{figure}\centering
\includegraphics[scale=0.6]{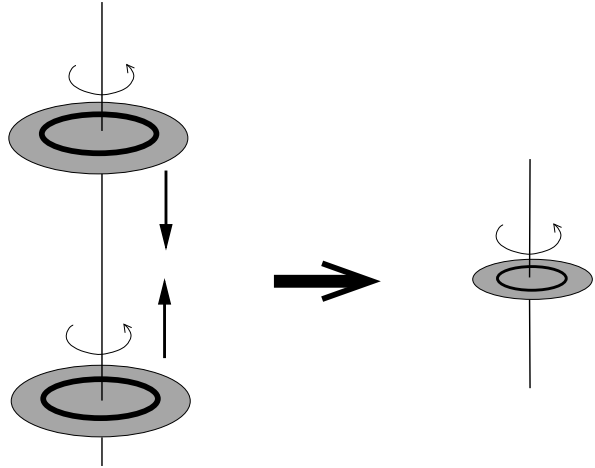}
%frueher: local_conservation.eps
\caption{Illustration of the local conservation equations.
Two corresponding rings of the initial
disks merge into a ring in the final disk. The
baryonic mass $\dd M_0$ and the angular momentum $\dd J$ of
these rings are conserved.}
\label{fig_local_conservation}
\end{figure}

We now turn to the main goal of this paper and analyse the formation of a
disk of dust under the influence of gravitational forces as the only
form of interaction.
Comparing the resulting \emph{differentially} rotating disk of dust with
the \emph{rigidly} rotating disk of the same baryonic mass $M_0$ and
angular momentum $J$ formed from the same initial disks
we may separate gravitational damping due to the emission of
gravitational waves from frictional processes in the matter.

We may interpret a disk of dust as a superposition of infinitesimally
thin dust rings.
Considering the geodesic motion of a single mass element,
one can show that for corresponding rings in the two initial disks
(see Fig.~\ref{fig_local_conservation})
the baryonic mass and the angular
momentum are conserved,
\begin{equation}\label{24}
\dd\tilde M_0=2\dd M,\quad \dd\tilde J=2\dd J,
\end{equation}
i.e. the baryonic masses $\dd M_0$ and the angular momenta
$\dd J$ of the rings with radius $\rho$, taking up the interval
[$\rho,\rho+\dd\rho]$,
in each of the two initial disks sum up to  
$\dd\tilde M_0=2\dd M_0$ and  $\dd\tilde J=2\dd J$
of the corresponding ring in the final disk (with radius $\tilde\rho$,
taking up the interval $[\tilde\rho,\tilde\rho+\dd\tilde\rho]$),
cf. Fig.~\ref{fig_local_conservation}.
It should be emphasized that the local conservation laws \eqref{24}
would be violated by dissipative processes in the matter or,
mathematically speaking, by dissipative terms in the total
energy-momentum tensor as the source of the Einstein equations during
the collision phase.
Having reached a final equilibrium configuration (e.g. a \emph{rigidly}
rotating disk of dust) the system \lq\lq forgets\rq\rq\ the dissipative
terms and behaves like cold dust with an energy momentum tensor of the
form \eqref{22}. During the interaction phase, angular momentum will be
transported within the disk by viscous forces and only the \emph{total}
angular momentum (axisymmetry!) and the total baryonic mass are
conserved \eqref{23}. The \emph{ring-wise} conservation of baryonic mass
and angular momentum \eqref{24} is characteristic for purely
gravitational damping processes. They arise from collision processes
governed by an energy-momentum tensor of dust without dissipative
terms. In this case the geodesic motion of the volume elements
implies the conservation of baryonic mass and angular momentum in each
volume element and therefore implies \eqref{24}. 

Eq.~\eqref{24} provides us with a subset of the boundary conditions
to be discussed in the next subsection.
It will turn out that these conditions, together with conditions
resulting from the field equations, determine a unique solution of the
Einstein equations describing a final disk with differential rotation
(DR disk) as the end product of the collision process.

\subsubsection{Boundary value problem for the final DR disk}
\label{BVP}

The line element \eqref{21}, which may also be used to describe
any axisymmetric and stationary \emph{differentially}
rotating  disk, can be reformulated to give
\begin{equation}\label{25}
\dd \tilde
s^2=\ee^{2\tilde\kappa}(\dd\tilde\rho^2+\dd\tilde\zeta^2)+\tilde\rho^2\ee^{-2\tilde\nu}(\dd\tilde\varphi-\tilde\omega\dd\tilde
t)^2-\ee^{2\tilde\nu}\dd \tilde t^2,
\end{equation}
where the usage of the functions $\tilde\kappa$, $\tilde\nu$ and $\tilde\omega$
[instead of $\tilde U$, $\tilde k$ and $\tilde a$ as in \eqref{21}] 
avoids numerical issues
with ergospheres (where $\ee^{2\tilde U}<0$). According to \eqref{22}
the energy-momentum
tensor is
\begin{equation}\label{25a}
\tilde T^{ij}=\tilde\varepsilon(\tilde\rho)\delta(\tilde\zeta)\tilde
u^i\tilde u^j
\end{equation}
and the four-velocity is again
[cf. \eqref{22a}] a linear combination of the killing
vectors, 
\begin{equation}\label{25b}
\tilde u^i=\ee^{-\tilde V}(\tilde\xi^i+\tilde\Omega\tilde\eta^i),
\end{equation}
where $\tilde V=\tilde V(\tilde\rho)$ and
$\tilde\Omega=\tilde\Omega(\tilde\rho)$ are functions of $\tilde\rho$
[constancy of $\tilde V$ and $\tilde\Omega$ defines rigid rotation,
cf.~\eqref{22a}].\footnote{All
quantities of the final DR disk are tilded.}

The vacuum field equations for $\tilde\nu$ and
$\tilde\omega$ are
%(cf. \cite{BlackHoles})
[cf. Bardeen (1973)]
\begin{equation}\label{26}
\bigtriangleup_1\tilde\nu
=\frac{\tilde\rho^2}{2}\ee^{-4\tilde\nu}
(\tilde\omega_{,\tilde\rho}^{\ 2}+\tilde\omega_{,\tilde\zeta}^{\ 2}),
\quad
\bigtriangleup_3\tilde\omega
=4(\tilde\nu_{,\tilde\rho}\tilde\omega_{,\tilde\rho}
+\tilde\nu_{,\tilde\zeta}\tilde\omega_{,\tilde\zeta}),
\end{equation}
with
\begin{equation}\label{27}
\bigtriangleup_n:=\partial_{\tilde\rho}^2+
\partial_{\tilde\zeta}^2+\frac{n}{\tilde\rho}\partial_{\tilde\rho}.
\end{equation}
The matter appears only in the boundary conditions along the disk
($\tilde\zeta=0$, $\tilde\rho<\tilde\rho_0$),
\begin{equation}\label{28}
\left.\tilde\nu_{,\tilde\zeta}\right|_{\tilde\zeta=0^+}
=2\pi\tilde\sigma\frac{1+\tilde v^2}{1-\tilde v^2},
\end{equation}
\begin{equation}\label{28a}
\left.\tilde\omega_{,\tilde\zeta}\right|_{\tilde\zeta=0^+}
=-8\pi\tilde\sigma\frac{\tilde\Omega-\tilde\omega}{1-\tilde v^2},
\end{equation}
\begin{equation}\label{29}
\tilde\rho(\tilde\Omega-\tilde\omega)^2=(1+\tilde
v^2)\ee^{4\tilde\nu}\tilde\nu_{,\tilde\rho}
+\tilde\rho^2(\tilde\Omega-\tilde\omega)\tilde\omega_{,\tilde\rho},
\end{equation}
where
\begin{equation}\label{30}
\tilde v:=\tilde\rho\ee^{-2\tilde\nu}(\tilde\Omega-\tilde\omega),\quad
\tilde\sigma:=\tilde\varepsilon\ee^{2\tilde\kappa}.
\end{equation}
Thus we have to deal with a boundary value problem for Einstein's vacuum
equations.

As already mentioned, the local conservation equations \eqref{24}
of the previous subsection lead to
additional boundary conditions along the disk. 
From $\dd\tilde M_0=2\dd M_0$ with $\dd M_0=2\pi\sigma\ee^{-V_0}\rho\dd\rho$
and $\dd\tilde M_0=2\pi\tilde\sigma\ee^{-\tilde V}\tilde\rho\dd\tilde\rho$
we obtain
\begin{equation}\label{31}
\tilde\sigma=2\sigma\frac{\rho\ee^{V_0}}{\tilde\rho\ee^{\tilde
V}}\frac{\dd\rho}{\dd \tilde\rho}.
\end{equation}
Likewise,  $\dd\tilde J=2\dd J$ with $\dd J=2\pi\sigma\ee^{-V_0}u^i\eta_i\rho\dd\rho$
and
  $\dd\tilde J
  =2\pi\tilde\sigma\ee^{-\tilde V}\tilde
  u^i\tilde\eta_i\tilde\rho\dd\tilde\rho$,
  $u^i\eta_i=\rho v\ee^{-V_0}$ and
  $\tilde u^i\tilde\eta_i=\tilde\rho\tilde v\ee^{-\tilde V}$
leads to
\begin{equation}\label{32a}
\tilde\rho\tilde v\ee^{-\tilde V}=\rho v\ee^{-V_0}.
\end{equation}
The function $\tilde V(\tilde\rho)$
can be calculated from $\tilde u^i\tilde u_i=-1$,
\begin{equation}\label{33}
\ee^{2\tilde V}=(1-\tilde v^2)\ee^{2\tilde\nu}.
\end{equation}
The remaining  boundary conditions describe the behaviour at infinity, where the
metric approaches the flat Minkowski metric,
\begin{equation}\label{33a}
\tilde\kappa=\tilde\nu=\tilde\omega=0,
\end{equation}
and in the plane $\tilde\zeta=0$ outside the disk
($\tilde\rho>\tilde\rho_0$), where \eqref{28} and
\eqref{28a} lead to vanishing normal derivatives,
\begin{equation}\label{33b}
\left.\tilde\nu_{,\tilde\zeta}\right|_{\tilde\zeta=0^+}=0,\quad
\left.\tilde\omega_{,\tilde\zeta}\right|_{\tilde\zeta=0^+}=0.
\end{equation}
In addition we have to ensure
regularity along the axis of symmetry $\tilde\rho=0$. 

Eqs.~\eqref{26}, \eqref{28}-\eqref{29} and \eqref{31}, \eqref{32a} form
a complete set of equations to determine the unknown
functions uniquely:
There are two two-dimensional functions, 
$\tilde\nu(\tilde\rho,\tilde\zeta)$ and
$\tilde\omega(\tilde\rho,\tilde\zeta)$, which have to satisfy the
two elliptic partial differential
equations \eqref{26} with the boundary conditions
\eqref{28} and \eqref{28a}, and three additional one-dimensional
functions in the disk, $\tilde\Omega(\tilde\rho)$,
$\tilde\sigma(\tilde\rho)$, $\rho(\tilde\rho)$, which have to obey
the three boundary conditions \eqref{29}, \eqref{31} and \eqref{32a}.  
(The metric function
$\tilde\kappa$ can be calculated by a line integral afterwards, but is
not needed for the computation of the efficiency $\eta$ in our
collision scenario.)

%%%%%%%%%%%%%%%%%%%%%%%%%%%%%%%
\subsubsection{Numerical method}

\begin{figure}\centering
\includegraphics[scale=1]{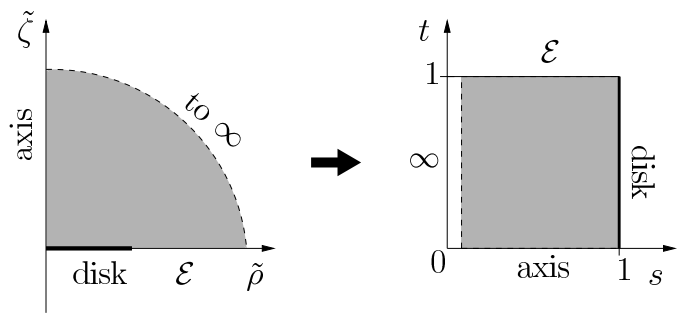}
\caption{The coordinate transformation \eqref{34}, \eqref{35} maps the
part $\tilde\zeta\ge 0$ of the $\tilde\rho$-$\tilde\zeta$-plane to
a unit square in the $s$-$t$-plane. $\mathcal E$ denotes
the equatorial plane outside the disk, $\tilde\zeta=0$,
$\tilde\rho>\tilde\rho_0$.}
\label{fig_Koordinatentransformation}
\end{figure}

In order to prepare numerical investigations
we will map the region $0\le \tilde\rho \le \infty$, $0\le \tilde\zeta\le
\infty$ to a unit square thus reaching a compactification of infinity,
cf. Fig.~\ref{fig_Koordinatentransformation}.
(Due to the reflection symmetry with respect
to the plane
$\tilde\zeta=0$ we can restrict ourselves to the region $\tilde\zeta\ge 0$.)
To do this we introduce in a first step elliptical coordinates
\begin{equation}\label{34}
\tilde \rho=\sqrt{(1+\xi^2)(1-\eta^2)},\quad \tilde\zeta=\xi\eta,\quad
\xi\in[0,\infty],\quad \eta\in[0,1]
\end{equation}
(without loss of generality,
we may choose units where $\tilde\rho_0=1$).
In a second step we stretch the coordinates by the transformation
\begin{equation}\label{35}
\xi=\cot\left(\frac{\pi}{2}s\right),\quad \eta=\sqrt{1-t},\quad
s\in[0,1], \quad t\in[0,1].
\end{equation}
The coordinates $s$ and $t$ form a unit square with the
following boundaries,
\begin{equation}
\begin{array}{ll}
s=0:  & \infty\\
s=1:  & \textrm{disk},\ \tilde\rho\le 1,\ \tilde\zeta=0\\
t=0:  & \textrm{axis of symmetry},\ \tilde\rho=0\\
t=1:  & \textrm{disk plane $\mathcal E$
                outside the matter},\ \tilde\rho>1,\ \tilde\zeta=0.
\end{array}
\nonumber
\end{equation}
The unknown functions in the boundary value problem are analytic
functions in this square (as is known for the case of Maclaurin disks or
the RR disks). Hence, it is convenient to use spectral methods for the
numerical solution of the boundary value problem.
We expand the unknown potentials in terms of
Chebyshev polynomials $T_j$ to a predetermined
order in the form
\begin{equation}\label{36}
f(s,t)=\sum\limits_{j,k}c_{jk} T_j(2s-1) T_k(2t-1)\quad \textrm{or}\quad
f(t)=\sum\limits_{k}c_k T_k(2t-1)\quad\textrm{(boundary)}
\end{equation}
and formulate the Einstein equations at the extrema of the Chebyshev
polynomials. This
leads to an algebraic system of equations for the Chebyshev coefficients (or,
alternatively, for the values of the potentials at these points) that
can be solved with the Newton-Raphson method. The iteration starts with an
initial \lq\lq guessed\rq\rq\ solution (for example the Newtonian
approximation, see Sec.~\ref{Newtonian} below).

\begin{figure}\centering
\includegraphics[scale=1.1]{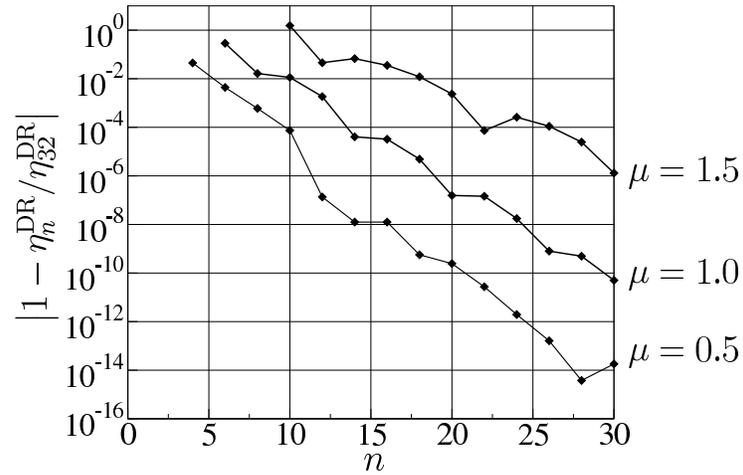}
\caption{Convergence properties of the numerical code for the example
$\eta^{\textrm{DR}}$. The values of the efficiency $\eta^{\textrm{DR}}$ for
different orders $n_s=n_t=n$ of the Chebyshev expansion
are related to the order $n=32$. The plot shows
$\left| 1-{\eta^{\textrm{DR}}_n}/{\eta^{\textrm{DR}}_{32}}\right|$ as
function of $n$.}
\label{Konvergenztest}
\end{figure}

The calculations show a decreasing accuracy of the numerical solution
for increasingly large values of the initial parameter $\mu$.
The reason are large
gradients of the metric potentials for strong relativistic DR disk which
make the Chebyshev approximation more costly.
To reach a better convergence we perform an additional coordinate
transformation
\begin{equation}\label{36a}
s=\frac{\sinh(\delta\cdot\tilde s)}{\sinh(\delta)}
\end{equation}
introducing a new coordinate $\tilde s$, where $\delta$ is a suitably
chosen parameter. As shown in
%\cite{Rings}
Ansorg~\&~Petroff (2005), this transformation smooths the
gradients of the metric functions.
The convergence is illustrated in Fig.~\ref{Konvergenztest}.

%%%%%%%%%%%%%%%%%%%%%%%%%%%
\subsubsection{Results}\label{results}

\begin{figure}\centering
\includegraphics[scale=0.3]{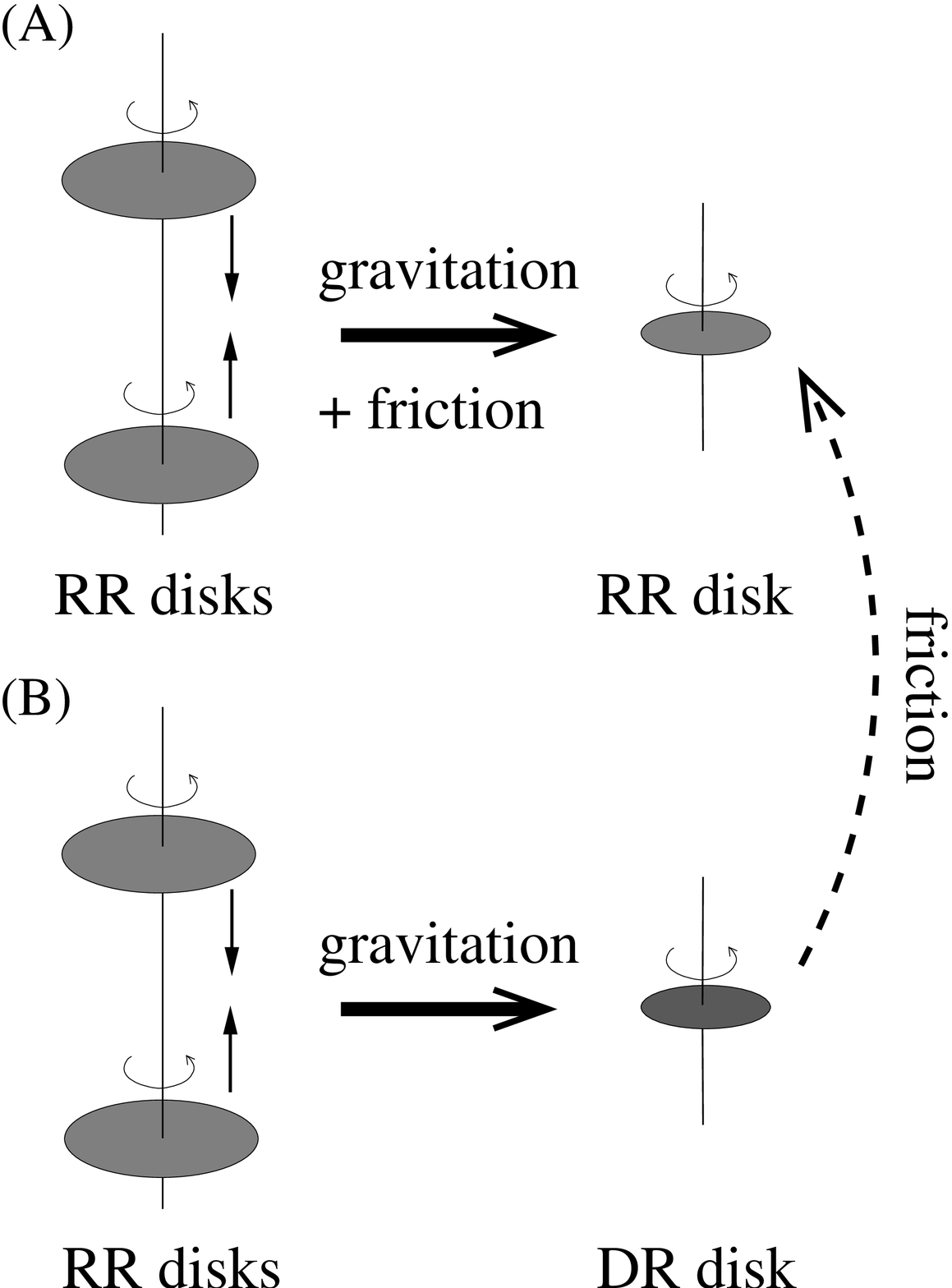} %frueher: model_Scheiben.eps
\caption{Two models for disk collisions:\newline
(A) Under the influence of
a small amount of friction, RR disks merge again into an RR disk. This
scenario was discussed in
%\cite{Hennig}
Hennig \& Neugebauer (2006), see Sec.~\ref{AssumptionRR}.
\newline
(B) In the absence of friction, the same RR disks merge into
a DR disk.
Allowing for friction afterwards, the system would again arrive at the RR
disk of scenario (A) after a sufficiently long time.
%\newline
%The baryonic mass and the angular momentum are conserved locally in
%scenario (B). However, in scenario (A), angular momentum will be
%transported within the disc by viscous forces.
}
\label{model_Scheiben}
\end{figure}

Using this numerical algorithm, we are able to solve the 
boundary value problem for the final DR disk.
In particular, we could calculate, for each value of the initial
parameter $\mu$,
all \emph{metric coefficients} of this final disk.
However, we will restrict ourselves to the discussion
of the relations between the initial and final \emph{parameters} and the
\emph{efficiency} of the collision process.
Especially, we will compare the
final DR disk with an RR disk having the same baryonic mass and
angular momentum.
The point made here is that such a rigidly rotating disk represents the
state of \lq\lq thermodynamic equilibrium\rq\rq\ for disks of dust as
the end point of their thermodynamic evolution.
As sketched in  Fig.~\ref{model_Scheiben},
there are at least two possibilities for the formation
of this final RR disk:
The direct process (A) including friction from the beginning
or the equivalent thermodynamic process (B) where, in a first step, a
\emph{differentially rotating} disk is formed (by gravitational damping
alone, no friction)
and, in a second step, the angular velocity becomes
\emph{constant} (due to friction).
Note that baryonic mass and angular momentum are conserved in both
processes. 
By comparing (A) and (B) we may
extract the contribution of friction in scenario~(A).

In the following discussion, tilded quantities, as before, belong to the final
DR disk of scenario (B) in Fig.~\ref{model_Scheiben},
a superscript \lq\lq RR\rq\rq\ denotes
quantities of the final RR disk in scenario (A) and the centrifugal
parameter $\mu$ without any additions characterizes the initial RR disks.  

\begin{figure}\centering
\includegraphics[scale=0.96]{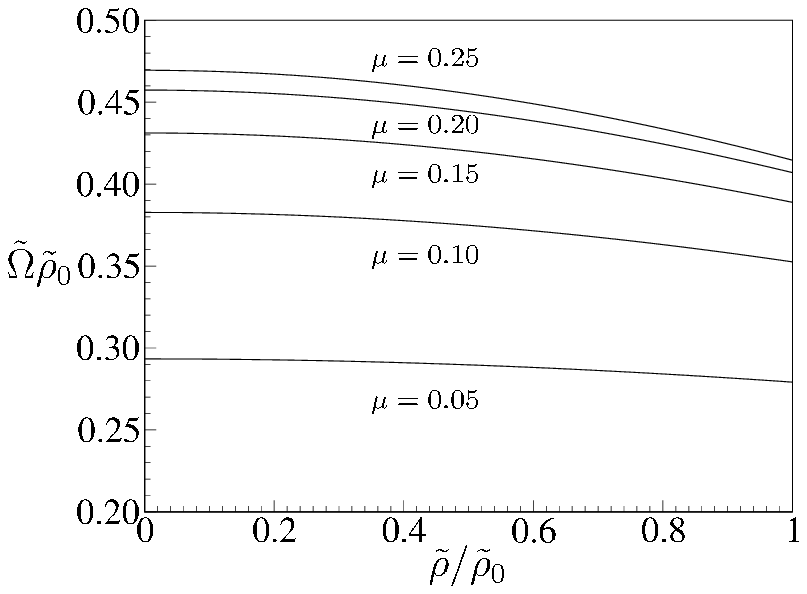}
\includegraphics[scale=0.96]{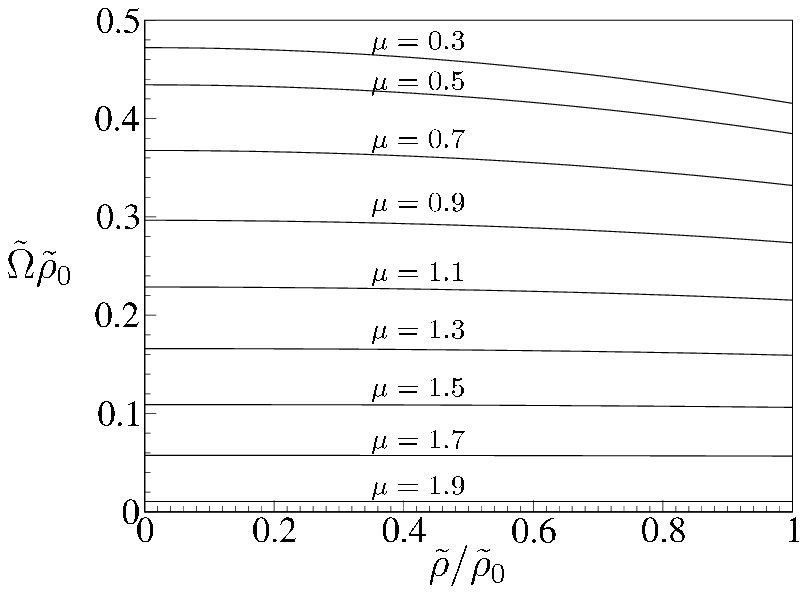}\\[2ex]
\includegraphics[scale=0.96]{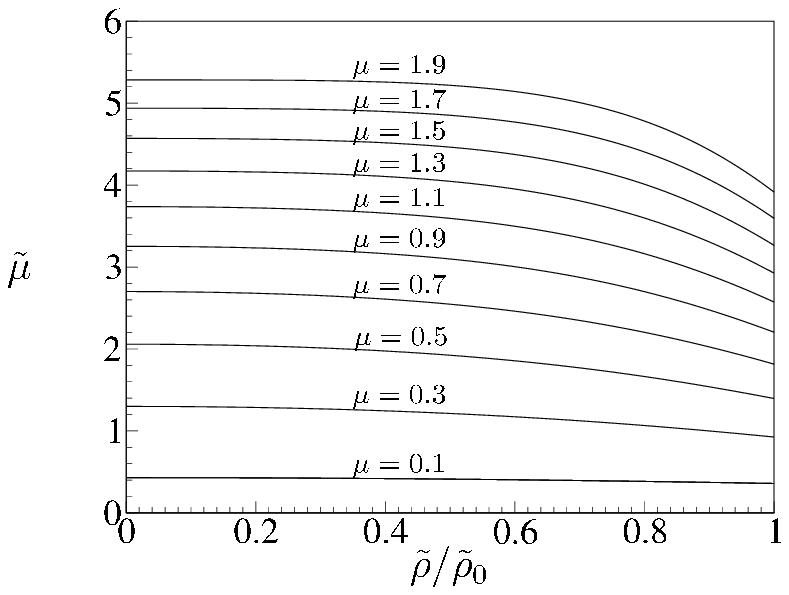}
\includegraphics[scale=0.96]{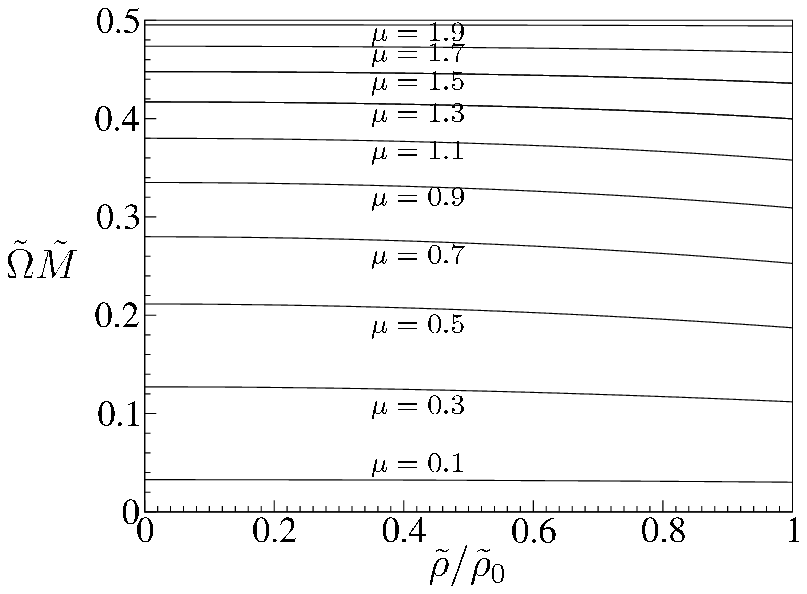}\\[2ex]
\includegraphics[scale=0.96]{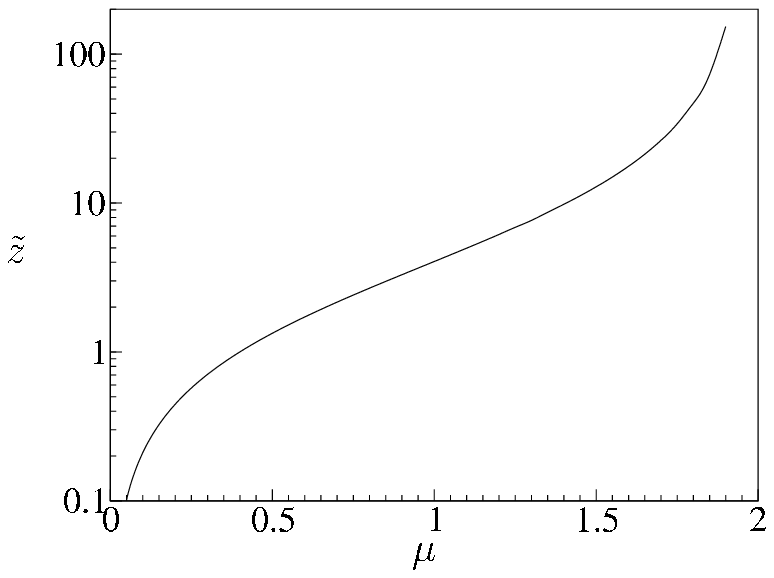}
\includegraphics[scale=0.96]{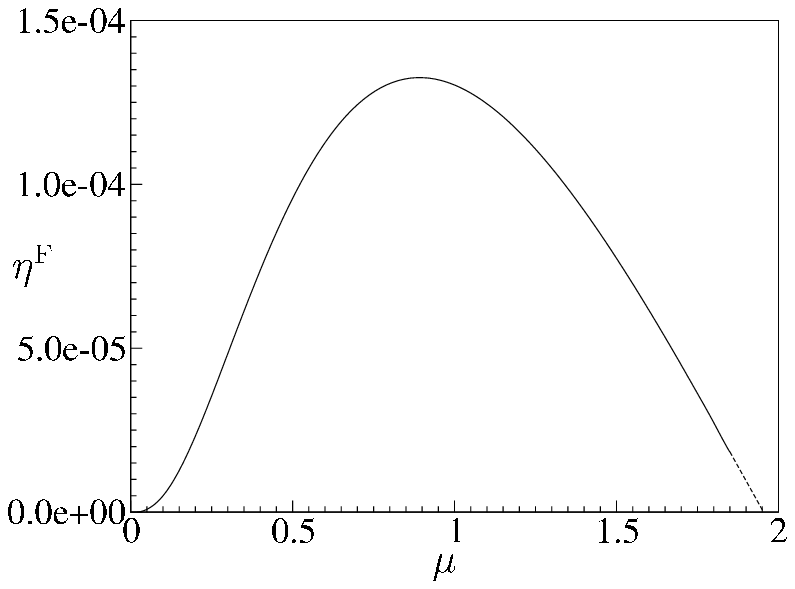}
\caption{Parameter relations for the collision of RR disks. We performed
numerical calculations for values of the initial centrifugal parameter
$\mu$ in the invervall $[0,1.9]$. The dotted part of the curve in the
last plot is an extrapolation for larger $\mu$. This extrapolation and
the rapidly growing redshift in the second last plot indicate, that the
initial parameter $\mu$ in scenario (B)
is limited (approximately, or perhaps even
exactly) to the same interval as in scenario (A), $0<\mu<1.954\dots$,
see Sec.~\ref{AssumptionRR}.}
\label{fig_DR}
\end{figure}

The rotation curve of the final DR disk, i.e. its
(normalized) angular velocity $\tilde\Omega\tilde\rho_0$
as a function of the (normalized) radius $\tilde\rho/\tilde\rho_0$
is shown in the first two plots of
Fig.~\ref{fig_DR}. For small parameters $\mu$ (post-Newtonian regime)
the function
$\tilde\Omega\tilde\rho_0$ is almost constant (first plot).
Interestingly, strongly relativistic disks ($\mu\gtrsim 1.5$)
show the same property
(second plot). Moreover,
$\tilde\Omega\tilde\rho_0$ tends to zero in the
ultrarelativistic limit in analogy to
the relation  $\Omega^{\textrm{RR}}\rho^{\textrm{RR}}_0\to 0$
which holds for RR disks in the
ultrarelativistic limit $\mu^{\textrm{RR}}\to 4.62966\dots$

The \lq\lq centrifugal parameter\rq\rq\
$\tilde\mu=2\tilde\Omega^2\tilde\rho_0^2\ee^{-\tilde V}=\mu(\tilde\rho)$ is
shown in the third plot of Fig.~\ref{fig_DR}. Like the angular
velocity, $\tilde\mu$ is almost constant for small $\mu$. For
strongly relativistic DR disks, $\tilde\mu$  in the center of the disk
exceeds the limit $\mu^{\textrm{RR}}_{\textrm{max}}=4.62966\dots$ of RR disks.

The fourth plot of Fig.~\ref{fig_DR} shows the quantity
$\tilde\Omega\tilde M$ as a function of $\rho/\rho_0$.
For strongly relativistic DR disks
$\tilde\Omega\tilde M$ becomes constant and approaches the limit $0.5$.
On the other hand, this is a characteristic value for extreme Kerr
black holes where
$\Omega_\textrm{H}M_\textrm{BH}=0.5$ ($\Omega_\textrm{H}$: angular
velocity of the horizon, $M_\textrm{BH}$: black hole mass). Indeed, one
can show that there is a phase transition between RR disks and Kerr
black holes \cite{Bardeen2, Neugebauer2}.
This inspires the conjecture that the DR disk exhibits the same
phase transition.
There is no obstacle for a (numerical) proof of this
assumption in principle.
To extend our present code to study the parametric collapse
of the DR disk including the formation of a horizon we would have
to follow the
ideas of Bardeen and Wagoner (1971)
%\cite{Bardeen2}
who analysed this problem
for the RR disks. However, such investigations are outside the scope of
this paper.

The fifth plot of  Fig.~\ref{fig_DR} shows the redshift $\tilde z$ for a
photon emitted from the disk center
as a function of the initial centrifugal parameter $\mu$.
For
increasing values of $\mu$ (relativistic DR disks) $\tilde z$ grows rapidly.

An important result is the efficiency $\eta^{\textrm{DR}}$ of the formation
of DR disks which measures the amount of energy converted into
gravitational radiation.
The difference $\eta^{\textrm{F}}=\eta^{\textrm{RR}}-\eta^{\textrm{DR}}$ as
shown in the last plot of Fig.~\ref{fig_DR} compares this value with
the efficiency $\eta^{\textrm{RR}}$ of the RR disk forming process as
sketched in scenario (A) of Fig.~\ref{model_Scheiben}.
Thereby, $\eta^{\textrm{F}}$ is the part of energy lost due to
friction during the formation of a final RR disk.
We find
$\eta^{\textrm{F}}=\eta^{\textrm{RR}}-\eta^{\textrm{DR}}<1.5\times 10^{-4}$,
i.e. the contribution of friction is extremely small,
$ \eta^{\textrm{F}}\ll\eta^{\textrm{RR}}$,
such that the gravitational radiation dominates the collision process (A).

%%%%%%%%%%%%%%%%%%%%%%%%%%%%%%%
\subsubsection{Analytical treatment of the Newtonian limit}
\label{Newtonian}

Our numerical investigations have shown that the angular velocity of 
the final DR disk
becomes closer and closer to a constant over the whole range of
$\tilde\rho/\tilde\rho_0$ as the centrifugal parameter $\mu$ tends to
zero (cf. the first plot of Fig.~\ref{fig_DR}).
This leads one to suspect that a final disk with a 
\emph{strictly} constant 
angular velocity will solve the boundary value problem 
as discussed in Sec.~\ref{BVP} in \emph{Newtonian}
theory. Interestingly, we can treat this problem analytically. This will
now be demonstrated. Strictly speaking, there is no gravitational
radiation in Newton's theory. However, this Newtonian
boundary value problem can be seen as the limit of a sequence of
relativistic collisions with decreasing $\mu$, all reaching a final
equilibrium state due to gravitational emission.
Moreover, the Newtonian solution can be used as a
starting point for the iterative calculation of the final relativistic
DR disk.

Since the Newtonian limit of the RR disk is the Maclaurin
disk  we have to study the collision of two identical Maclaurin
disks using the local conservation laws~\eqref{24}.
The Newtonian potential $\tilde U$ of the final disk is a
solution of the Poisson equation
\begin{equation}\label{N1}
\bigtriangleup\tilde U = 4\pi\tilde\sigma\delta(\zeta)
\end{equation} 
with the boundary condition
\begin{equation}\label{N2}
\left.\tilde U_{,\tilde\zeta}\right|_{\tilde\zeta=0+}=2\pi\tilde\sigma,
\end{equation}
where $\tilde\sigma=\tilde\sigma(\tilde\rho)$ is the surface mass
density of the final disk. With $\dd M=2\pi\sigma\dd\rho$ and $\dd
J=\Omega\rho^2\dd M$, Eq.~\eqref{24} leads to the additional boundary
conditions
\begin{equation}\label{N3}
\tilde\sigma(\tilde\rho)=
2\sigma(\rho)\frac{\rho}{\tilde\rho}\frac{\dd\rho}{\dd\tilde\rho},\quad
\tilde\Omega(\tilde\rho)\tilde\rho^2=\Omega_0\rho^2.
\end{equation}
The initial surface mass density of the Maclaurin disk is
\begin{equation}\label{N4}
\sigma(\rho)=\frac{3M}{2\pi\rho_0^2}\sqrt{1-\frac{\rho^2}{\rho_0^2}}
\end{equation}
and the initial constant angular velocity $\Omega_0$ is related to the
initial mass by
\begin{equation}\label{N5}
\Omega_0^2=\frac{3\pi M}{4\rho_0^3}.
\end{equation}
Using these relations, together with the Euler equation
\begin{equation}
\left.\tilde U_{,\tilde\rho}\right|_{\tilde\zeta=0}=\tilde\Omega^2(\tilde\rho)
\tilde\rho,
\end{equation}
we find that a (rigidly rotating) Maclaurin disk with the parameters
\begin{equation}\label{N6}
\tilde\Omega=4\Omega_0,\quad \tilde\rho_0=\frac{1}{2}\rho_0
\end{equation}
indeed solves the boundary value problem \eqref{N1}-\eqref{N3}.

%%%%%%%%%%%%%%%%%%%%%%%%%%%%%%%%%%%%%%%%%%%%%%%%%%%%%%%%%%%%%%%%%%%%%%%%%%%%%
\section{Discussion}

In this paper we have performed the analysis
of collision processes in the spirit of equilibrium thermodynamics.
Avoiding the solution 
of the
full dynamical problem, we compared initial and final
equilibrium configurations to obtain a \lq\lq rough\rq\rq\
picture of these processes. In this way we were able to calculate the
energy loss by the emission of gravitational waves and to find conditions
(\lq\lq parameter relations\rq\rq) for the formation of final stars and disks.

The application of this method to
collisions of perfect fluid stars and collisions of
rigidly rotating disks of dust leads to restrictions of the
initial parameters.
It turned out that the formation of final stars/disks from stars/disks
is only possible
for a subset of the parameter space of the
initial objects. Otherwise, the collision of spheres and disks
would lead to other final
states, e.g. to black holes.

Our main result is the numerical solution of the Einstein equations for the
differentially rotating (DR) disk formed
by the collision of two identical rigidly rotating (RR)
disks with parallel angular
momenta. We calculated the characteristic quantities of the final
DR disk, as for
example the rotation curve $\tilde\Omega(\tilde\rho)$ as it depends on
the centrifugal
parameter $\mu$ of the initial RR disks. It turned out, that the angular
velocity $\tilde\Omega$
is almost constant (as shown in Sec.~\ref{Newtonian}, it is
\emph{strictly} constant in the Newtonian limit). Therefore, the simplified
model of the formation of an RR disk from the
collision of two RR
disks as presented in
%\cite{Hennig}
Hennig \& Neugebauer (2006),
which has to allow frictional processes to reach constant angular velocity,
turns out to be a good approximation to our present purely gravitational
(frictionless) model (B).

\begin{table}[t]\centering
\begin{tabular}{lr}
\hline
{\it colliding objects} & $\eta_\textrm{max}$\\
\hline\hline
Schwarzschild BHs         & 29.3\%\\
RR disks                  & 23.8\%\\
Schwarzschild stars \quad & 19.7\%\\
Neutron stars             & 2.3\%\\
\hline
\end{tabular}
\caption{Upper limits for the efficiency $\eta$ of different collision
processes including Hawking's and Ellis' limit for the collision of two
spherically symmetric black holes \cite{Hawking}.
According to the last plot of Fig.~\ref{fig_DR}
[$\eta^{\textrm{F}}(1.954\dots)=0$] the two efficiencies
$\eta^{\textrm{RR}}$ and $\eta^{\textrm{DR}}$ coincide with a maximum
value
$\eta^{\textrm{RR}}_{\textrm{max}}=\eta^{\textrm{DR}}_{\textrm{max}}\approx
23.8\%$. 
}
\label{table}
\end{table}

For each of the studied collision scenarios, we calculated an upper limit
for the energy of the emitted gravitational waves. A summary of the maximal
efficiencies is given in table~\ref{table}. The value
$\eta_{\textrm{max}}\approx 2.3\%$ for the collision of Neutron stars is
relatively small compared to the other examples. The reason is the
restricted equation of state (completely degenerate ideal Fermi gas)
that does not allow for strongly relativistic stars.

%%%%%%%%%%%%%%%%%%%%%%%%%%%%%%%%%%%%%%%%%%%%%%%%%%%%%%%%%%%%%%%%%%%%%%%%%%%%%
\begin{acknowledgments}
We would like to thank David Petroff for many valuable discussions.
This work was supported by the Deutsche Forschungsgemeinschaft (DFG)
through the SFB/TR7 \lq\lq Gravitationswellenastronomie\rq\rq.
\end{acknowledgments}

%%%%%%%%%%%%%%%%%%%%%%%%%%%%%%%%%%%%%%%%%%%%%%%%%%%%%%%%%%%%%%%%%%%%%%%%%%%%
\appendix

\section{Potentials of the rigidly rotating disk of dust}\label{AppendixA}

For the numerical calculation of the DR disk that is formed by the
collision of two RR disks we need some formulae for
quantities of the RR disk of dust.

The coefficient $V_0$ in the four-velocity \eqref{22a} as a
function of the parameter $\mu$ can be calculated from a
very rapidly converging
series, cf. \cite{Kleinwaechter},
\begin{eqnarray}\label{A1}
\coth\frac{V_0}{2} & = & -\frac{4}{\mu}+
0.0294938052100425142\mu+5.4681333461446\cdot10^{-6}\mu^3\nonumber\\
&& -1.07467432587\cdot10^{-9}\mu^5+2.1127368\cdot10^{-13}\mu^7\\
&& -4.154\cdot10^{-17}\mu^9+\mathcal O(\mu^{11}).\nonumber
\end{eqnarray}
The disk values ($\zeta=0$, $\rho\le\rho_0$)
of the metric functions $U$ and $a$ and the mass density $\sigma$ are given
by the equations
\begin{equation}\label{A2}
\ee^{2U}=\ee^{2V_0(\hat\mu)}-\frac{\mu\rho^2}{2\rho_0^2},
\end{equation}
\begin{equation}\label{A3}
(1+\Omega_0 a)\ee^{2U}=\ee^{V_0(\mu)}\ee^{V_0(\hat\mu)},
\end{equation}
\begin{equation}\label{A4}
\sigma=-\frac{\Omega_0}{2\pi\ee^{V_0(\mu)}}
\frac{b_0'(\hat\mu)}
{\ee^{V_0\hat\mu}},
\end{equation}
with
\begin{equation}\label{A5}
\Omega_0=\sqrt{\frac{\mu}{2}}\frac{\ee^{V_0}}{\rho_0},\quad
b_0=-\sqrt{1-\ee^{4V_0}-4\Omega_0^2\rho_0^2},
\end{equation}
cf. \cite{Neugebauer2}.
The notation $V_0(\hat\mu)$, $b'_0(\hat\mu)$ indicates that the argument
$\mu$ in the parameter functions $V_0(\mu)$ and $b'_0(\mu)$ has to be
replaced by $\hat\mu=(1-\rho^2/\rho_0^2)\mu$. $b'_0(\hat\mu)$ means
$\dd b_0(\hat\mu)/\dd\hat\mu$.

%%%%%%%%%%%%%%%%%%%%%%%%%%%%%%%%%%%%%%%%%%%%%%%%%%%%%%%%%%%%%%%%%%%%%%%%%%%%%

\end{document}